\documentclass[11pt]{article}

\usepackage{amssymb, amsmath}
\usepackage{graphicx}
\usepackage{cite}

\newcommand{\const}{\,{\rm const}\,}

\def\be{\begin{equation}}
\def\ee{\end{equation}}
\def\bea{\begin{eqnarray}}
\def\eea{\end{eqnarray}}

\title{\bf{Some Gravitational Instantons}}

\author{Don N. Page
\thanks{This is a paper I presented at a conference in Moscow in
December 1978, but it was never published.  Hari K. Kunduri kindly typed
it up for those interested in its description of the four-dimensional
gravitational instantons with positive cosmological constant known in
1978.}
\\ \\
\small \sl University of Cambridge \\
\small \sl Department of Applied Mathematics and Theoretical Physics \\
\small \sl Silver Street 
\small \sl Cambridge, CB3 9EW, England} 
\date{}
\begin{document}

\maketitle

\begin{abstract}

All known gravitational instantons with $\Lambda > 0$ ($S^4, CP^2, S^2 \times S^2$ and $CP \# \overline{CP}^2$) are described. They are all special cases of the Taub-NUT-$\Lambda$ local solution. Hence they are Type D and are locally conformably K\"ahler. They all may be expressed in Bianchi IX form and have four or more Killing vectors.
\end{abstract}

\section{Introduction}
In Euclidean Quantum Gravity\cite{Page1}, one calculates amplitudes as path integrals over various classes of positive definite metrics $g_{ab}$:
\begin{equation}
Z = \int D[g_{ab}] e^{-\hat{I}[g_{ab}]}
\end{equation} where the action may be taken as
\begin{equation}
\hat{I}[g_{ab}] = -\frac{1}{16\pi}\int_{M} (R - 2\Lambda)\sqrt{g}d^4x - \frac{1}{8\pi}\int_{\partial{M}} K \sqrt{h} d^3 x
\end{equation} The dominant contribution is expected to come from instantons or extrema of the action. These are four-dimensional complete regular manifolds with positive-definite metrics that are solutions of the classical Einstein equations
\begin{equation}
R_{ab} = \Lambda g_{ab}
\end{equation} \par There are at least three kinds of gravitational instantons, each of which will be important in different physical processes. First, there are instantons which are asymptotically flat in all directions. A large class of examples is known\cite{Page2,Page3} which at large distances approach the flat metric on $R^4$ with a cyclic angular identification. These instantons may contribute to various transition amplitudes or S-matrix elements. \par Second, there are instantons which ar e periodic in some imaginary time variable but asymptotically flat in spatial directions. The best known example is the Euclidean Schwarzchild solution\cite{Page4}, but there are others such as the multi-Taub-NUT metrics \cite{Page5} in which the time direction has a certain twist. These instantons contribute to the partition function of the thermal canonical ensemble of the gravitational field. \par Third, there are compact instantons, which may be considered as approximations to spacetime foam\cite{Page6}, or the structure of spacetime at very small distances. Unlike the other kinds, these instantons may have nonzero $\Lambda$ as a Lagrange multiplier in the volume canonical ensemble
\begin{equation}
Z[\Lambda] = \sum_{n} \langle g_n \mid e^{-\frac{\Lambda V}{8\pi}}\mid g_n \rangle
\end{equation} from which the density of states may be obtained by an inverse Laplace transform\cite{Page6}:
\begin{equation}
N(V) = \frac{1}{16\pi^2 i}\int_{-i\infty}^{i\infty} d\Lambda Z[\Lambda]e^{\frac{\Lambda V}{8\pi}}.
\end{equation} Most of these compact instantons will have $\Lambda <0$ and be very complicated, but for simplicity I will consider only those with $\Lambda > 0$.
\section{Metrics for known instantons with $\Lambda >0$} Unlike other cases in which infinite families of instantons are known, the restriction to $\Lambda >0$ leaves only four known examples (up to orientations and identifications): the Einstein metrics on the manifolds $S^4, CP^2, S^2 \times S^2$, and $CP^2 \# \overline{CP}^2$. The first three are homogenous spaces long known to geometers, but the fourth was discovered recently \cite{Page7}.  \par The metrics of these instantons are all special cases of the Taub-NUT-$\Lambda$ metric
\begin{equation}
ds^2 = (1- \frac{4}{3}\Lambda N^2)^{-1}[P^{-1}dr^2 + 4N^2P(d\psi + \cos\theta d\phi)^2 + (r^2-N^2)(d\theta^2 + \sin^2\theta d\phi^2)]
\end{equation} where
\begin{equation}
P = \frac{r^2 - 2Mr + N^2 - \frac{1}{3}\Lambda (r^2-N^2)^2}{(1-\frac{4}{3}\Lambda N^2)(r^2-N^2)}
\end{equation} This in turn is a special case of a six-parameter electrovac solution discovered by Carter \cite{Page8} which has been further generalized by Plebanski and Demianski \cite{Page9} to seven parameters, apparently the most general Type D local solution to the Einstein-Maxwell equations with a $\Lambda$ term. However, the general solution is not regular globally, having various singularities for arbitrary values of the parameters. Even metric (6) is generally singular at the zeroes and poles of $P(r)$. But for certain values of the parameters, the singularities become mere coordinate singularities, like the axis in polar coordinates. Then the manifold becomes regular everywhere, a gravitational instanton. \par The first special case is the Einstein metric on $S^4$, the four-sphere. This can be obtained from (6) by setting $\tau = 2N\psi$ and taking the limit $M=N \to 0$:
\begin{equation}
ds^2 = (1- \frac{1}{3}\Lambda r^2)^{-1}dr^2 + (1-\frac{1}{3}\Lambda r^2)d\tau^2 + r^2\left(d\theta^2 +\sin^2\theta d\phi^2\right).
\end{equation} This is the analytic continuation to the Euclidean section $\tau = it$ of the de Sitter metric \cite{Page10}. The apparent singularity at $r= r_0 = (3/\Lambda)^{\frac{1}{2}}$ can be removed by making $\tau$ periodic with period $2\pi r_0$; i.e., let $r= r_0\cos\chi$ and $\tau= r_0 \eta$ so that the metric becomes
\begin{equation}
ds^2 = 3\Lambda^{-1}[d\chi^2 + \sin^2\chi d\eta^2 + \cos^2\chi \left(d\theta^2 + \sin^2\theta d\phi^2\right)].
\end{equation} Then $\chi = 0$ is an axis of spherical polar coordinates $(\chi,\eta)$ and the manifold is perfectly regular there. This axis is a 2-sphere in the coordinates $(\theta,\phi)$, a `bolt' in the language of \cite{Page11}. \par The second special case is $CP^2$, the complex projective two space, which is the space of the equivalence classes of complex number triples up to an overall multiplication. The Einstein metric may be obtained from (6) by setting $\Lambda = 0.75 r_0^{-2}$, $N=r_0 - \epsilon$, $r= r_0 - \epsilon \cos\chi$ and taking the limit $\epsilon \to 0$:
\begin{equation}
ds^2 = \frac{3}{2}\Lambda^{-1}\left[d\chi^2 + \frac{1}{4}\sin^2\chi(d\psi + cos\theta d\phi)^2 + \sin^2\frac{\chi}{2}(d\theta^2 + \sin^2\theta d\phi^2)\right].
\end{equation} Here the surfaces $\chi = \const$ are topologically 3-spheres  $(S^3)$  labeled by Euler angles $(\psi,\theta,\phi)$, so $\psi$ has period $4\pi$. Hence $\chi=0,\pi$ are regular axes of polar coordinates $(\sin\chi,\psi/2)$. The axis $\chi=\pi$ is a 2-sphere or bolt as is $\chi=0$ in $S^4$, but the axis $\chi=0$ is as single point or `nut' \cite{Page11}. Although the 2-sphere $(\theta,\phi)$ shrinks to zero at the nut, the manifold is perfectly regular there, since it is simply the origin of hyperspherical coordinates $(\chi,\psi,\theta,\phi)$. The properties of $CP^2$ as a gravitational instanton have been discussed in \cite{Page12} and \cite{Page13}.
\par The third special case is $S^2 \times S^2$, the product of the two-2-spheres. The Einstein metric may be obtained setting $\Lambda = (r_0^2 + \frac{1}{3}\epsilon^2)^{-1}$, $M=\frac{2}{3}r_0(r_0^2 - \epsilon^2)\Lambda$, $r= r_0 - \epsilon \cos\chi$, $\psi = \frac{r_0^2}{2N\epsilon} \eta$ and taking the limit $N \to 0, \epsilon \to 0$ in (6):
\begin{equation}
ds^2 = \Lambda^{-1}\left[d\chi^2 + \sin^2\chi d\eta^2 + d\theta^2 + \sin^2\theta d\phi^2\right].
\end{equation} Here both $\chi =0$ and $\chi = \pi$ are bolts, the north and south poles of one $S^2$ cross the entire other $S^2$.
\par The fourth special case is $CP^2 \# \overline{CP}^2$ the complex projective two space attached to its complex conjugate by removing a 4-ball from each manifold and gluing together the resulting $S^3$ boundaries. The Einstein metric may be obtained \cite{Page7} from (6) by setting $M=0, N=(3+3\nu^2)^{1/2}(1-\nu^2)^{-1}\Lambda^{-1/2}$, and $\nu^4 + 4\nu^3 - 6\nu^2 + 12\nu - 3 =0$:
\begin{equation}
ds^2 = \left(\frac{3 + 3\nu^2}{3 + \nu^2}\right)\Lambda^{-1}\left[U^{-1}d\chi^2 + \frac{1}{4}U\sin^2\chi(d\psi + \cos\theta d\phi)^2 + \frac{1}{4}(\nu^{-1} - \nu\cos^2\chi)(d\theta^2 + \sin^2\theta d\phi^2)\right]
\end{equation} where
\begin{eqnarray}
U &=& 1 - \frac{2\nu^2\sin^2\chi}{(3+\nu^2)(1-\nu^2\cos^2\chi)} \\
\nu &=& -1 - (2 + a - a^{-1})^{1/2} + [4 - a + a^{-1} + 8(a-a^{-1})^{-1/2}(a+ a^{-1})^{-1}] \sim 0.28170156,  \\
a &=& (2^{1/2} + 1)^{1/3}
\end{eqnarray} As in the $CP^2$ metric, $(\psi,\theta,\phi)$ are Euler angles on the 3-spheres of constant $\chi$, but the $CP^2$ nut at $\chi =0$ has been replaced by the bolt of $\overline{CP}^2$, i.e., the point $\chi=0$ has been `blown up' \cite{Page14}.
\section{Properties of these instantons} The different topologies of these instantons can be described in terms of their topological quantum numbers:
\begin{eqnarray}
\textrm{Euler Number} \phantom{0} \chi &=& \frac{1}{128\pi^2}\int R_{abcd}R_{efgh}\epsilon^{abef}\epsilon^{cdgh}\sqrt{g}d^4x. \\
\textrm{Signature} \phantom{0} \tau &=& \frac{1}{96\pi^2}\int R_{abcd}R^{ab}_{\phantom{gh}gh}\epsilon^{cdgh}\sqrt{g}d^4x.
\end{eqnarray} These are listed below:
\begin{table}[ht]
\centering
\begin{tabular}{c c c c c}
\hline
\phantom{Type} & $S^4$ & $CP^2$ & $S^2 \times S^2$ &$CP^2 \# \overline{CP}^2$ \\
$\chi$ & 2 & 3 & 4 & 4\\
$\tau$ & 0 & 1 & 0 & 0 \\
\hline
\end{tabular}
\end{table}

\par For compact simply-connected 4-manifolds with spin structure, $\chi$ and $\tau$ re believed to classify the topology completely \cite{Page6}. However, $CP^2$ and $CP^2 \# \overline{CP}^2$ do not admit spinors consistently but only generalized spin structures \cite{Page15}. On needs the intersection matrix of the 2-cycles (the independent 2-surfaces that cannot be shrunk to a point) to classify completely the homotopy type of a simply-connected compact instanton with $\Lambda >0$. This limits such instantons to being homotopic to $S^4$, the confected sum o $n$ copies of $S^2 \times S^2$, or the connected sum of $m$ copies of $CP^2$ and $n$ of $CP^2$ with $3|m-n| < 2(m+n+2)$ \cite{Page16}. The instantons given above are the only ones known.  \par $S^4$ has no 2-cycles and hence a vacuous intersection matrix. $CP^2$ has one 2-cycle, the bolt $\chi=\pi$ with self-intersection number 1 (i.e., this 2-surface intersects a slightly displaced version of itself once). $S^2 \times S^2$ has two 2-cycles , $(\chi=\chi_0, \eta = \eta_0)$ and $\theta=\theta_0, \phi = \phi_0$ with self-intersection numbers 0 and 0. $CP^2 \# \overline{CP}^2$ also has two 2-cycles, $\chi=0$ and $\chi = \pi$ but they have self-intersection numbers $+1$ and $-1$. Only manifolds whit even self-intersection numbers have spin structures.
\par Despite their topological differences, the known instantons with $\Lambda >0$ have a striking number of characteristics in common. As shown above, they call all be obtained from the Taub-NUT-$\Lambda$ metric (6) and so are Type D \cite{Page9}. This can be shown to imply that they ar conformably K\"ahler metrics. I.e., the metric may be conformably transformed, $g_{ab} \to \Omega^2 g_{ab}$, so that the self-dual or anti-self dual form $F_{ab} \pm \frac{1}{2}\epsilon_{abcd}F^{cd}$ that exists on it becomes covariantly constant and locally
\begin{equation}
\Omega^2 g_{ab}dx^a dx^b = K_{,\alpha \bar{\beta}}dz^{\alpha}dz^{\bar{\beta}}
\end{equation} for some scalar $K$ and complex coordinates $z^1 = \bar{z}^1$ and $z^2 = \bar{z}^2$. The conformal factor is $\Omega = (r \pm N)^{-1}$. For $S^4$ it is singular and turns the matric into the Bertotti-Robinson metric. The $CP^2$ and $S^2 \times S^2$ metrics already K\"ahler , so the conformal factor is constant for them. However, for $CP^2 \# \overline{CP}^2$ the conformal factor is regular but not constant. Thus this last instanton is a counterexample to the conjecture that all Einstein metrics on K\"ahler manifolds are K\"ahler \cite{Page16}.
\par Another interesting similarity is that all these instantons can be put in Bianchi IX form: Let
\begin{equation} \omega_1 = \frac{1}{2}(d\psi + \cos\theta d\phi),\phantom{0} \omega_2 = \frac{1}{2}(\cos\psi d\theta + \sin\psi \sin\theta d\phi),\phantom{0} \omega_3 = \frac{1}{2}(\sin\psi d\theta - \cos\psi \sin\theta d\phi)
\end{equation} be the one-forms that are left invariant under the $SU(2)$ isometry group acting transitively on the 3-surfaces labeled by $\chi = \const$. They obey the structure equations
\begin{equation}
d\psi_i = \epsilon_{ijk}\omega_j \wedge \omega_k
\end{equation} where $\epsilon_{ijk}$ is the 3-dimensional anti-symmetric tensor. Then the general Bianchi IX metric \cite{Page17}, cast in positive-definite form, is
\begin{equation}
ds^2 = f^2(d\chi^2 + a^2\omega_1^2 + b^2\omega_2^2 + c^2\omega_3^2),
\end{equation} where $f,a,b,c$ are all functions of $\chi$. A 3-space with $fa = fb = fc = 1$ has the unit 3-sphere metric
\begin{equation}
ds^2 = \omega_1^ 2 + \omega_2^2 + \omega_3^2 = d\left(\frac{\theta}{2}\right)^2 + \cos^2\frac{\theta}{2} d\left(\frac{\psi + \phi}{2}\right)^2 + \sin^2\frac{\theta}{2}d\left(\frac{\psi-\phi}{2}\right)^2.
\end{equation} By changing variables as necessary , the instanton metrics above may be put in the form
\begin{eqnarray}
S^4: ds^2 &=& 3\Lambda^{-1}\left[d\chi^2 + \sin^2\chi(\omega_1^2 + \omega_2^2 + \omega_3^2)\right], \\
CP^2: ds^2 &=& \frac{3}{2}\Lambda^{-1}\left[d\chi^2 + \sin^2\chi \omega_1^2 + 4\sin^2\frac{\chi}{2}(\omega_2^2 + \omega_3^2)\right], \\
S^2 \times S^2: ds^2 &=& \frac{1}{2}\Lambda^{-1}\left[
d\chi^2 + 16\sin^2\frac{\chi}{2}\omega_1^1 + 16\cos^2\frac{\chi}{2}\omega_2^2 + 16\omega_3^2\right], \\
CP^2 \# \overline{CP}^2: ds^2 &=& \left(\frac{3 + 3\nu^2}{3+\nu^2}\right)\Lambda^{-1}\left[U^{-1}d\chi^2 + U\sin^2\chi \omega_1^2 + (\nu^{-1} - \nu\cos^2\chi)(\omega_2^2 + \omega_3^2)\right],
\end{eqnarray} where $\nu$ and $U$ are given by Eqs. (13-15). In each case $\chi$ varies from $o$ to $\pi$, which are regular axes of polar coordinates where $a, b$, or $c$, or all three, vanish as $n\chi$, if the corresponding one-form(s) $\omega_i$ with vanishing coefficient has period $2\pi/n$. For $S^4, CP^2$, and $CP^2 \# \overline{CP}^2$, $n=1$ and so $(\psi,\theta,\phi)$ are Euler angles on $S^3$ with the identification $(\psi,\theta,phi) = (\psi + 2\pi j, \theta,\phi + 2\pi k)$ for all pairs of integers $j,k$ either both even or both odd. For $S^2 \times S^2$, the Bianchi IX form above was found by Chris Pope and has $n=2$, so antipodal points on the 3-sphere must be identified. I.e., the $\chi=\const$ surfaces are $RP^3$'s with Euler angles $(\psi,\theta,\phi)$ having the identification $(\psi,\theta,\phi) = (\psi + 2\pi j, \theta, \phi+ 2\pi k)$ for al integers $j,k$, so that each $\omega_i$ has period $\pi$ instead of $2\pi$. The coordinates on the two $S^2$'s back in the metric form (11) (where $\theta,\phi$ are not the same variables as in (19)) are those of the directions of the blades of a pair of scissors which has been opened by an angle $\chi$ and then rotated by the Euler angles $(\psi,\theta,\phi)$.
\par Since the instanton metrics can all be put in Bianchi IX form, they each have at least three Killing vectors, corresponding to the $SU(2)$ isometries. Actually, they all have more: $S^4$ has 10 (since it is maximally symmetric), $CP^2$ has 8 \cite{Page13}, $S^2 \times S^2$ has 5 (3 on each $S^2$), and $CP^2 \# \overline{CP}^2$ has 4 $ (\partial/\partial \psi$ as well as those generating the $SU(2)$ isometry).
\par One might conjecture that all gravitational instantons with $\Lambda > 0$ must share several of these characteristics. However, these properties may simply be selection effects that allowed these instantons to be found. The next simplest instanton, if it exists, would probably be an Einstein metric on $CP^2 \# CP^2$. It cannot have Bianchi IX form and can be shown not to lie within the Type D metrics. Whether it or or more complicated possibilities exist remain to be seen.

\end{document}